# Orphan optical flare as SOSS emission afterglow, localization in time

V. Lipunov,[1,2]★ V. Kornilov,[1] K. Zhirkov,[1] N. Tyurina,[2] E. Gorbovskoy,[2] D. Vlasenko,[1] S. Simakov,[2]
V. Topolev,[1] C. Francile,[3,4] R. Podesta,[3,4] F. Podesta,[3,4] D. Svinkin,[5] N. Budnev,[6] O. Gress,[2,6]
P. Balanutsa,[2] A. Kuznetsov,[2] A. Chasovnikov,[1] M. Serra-Ricart,[7] A. Gabovich,[2,8] E. Minkina,[1,2]
G. Antipov,[2] S. Svertilov,[1,9] A. Tlatov,[10] V. Senik,[2] Yu. Tselik,[2] Ya. Kechin[2] and V. Yurkov[8]

[1] *Physics Department, M. V. Lomonosov Moscow State University, Leninskie gory, GSP-1, Moscow 119991, Russia*
[2] *M. V. Lomonosov Moscow State University, SAI, Universitetsky pr., 13, Moscow 119234, Russia*
[3] *San Juan National University, Casilla de Correo 49, San Juan 5400, Argentina*
[4] *Observatorio Astronomico Felix Aguilar (OAFA), Avda Benavides 8175, Rivadavia, El Leonsito, San Juan 5400, Argentina*
[5] *Ioffe Institute, 6 Politekhnicheskaya, St Petersburg 194021, Russia*
[6] *Irkutsk State University, Applied Physics Institute, 20, Gagarin blvd, Irkutsk 664003,, Russia*
[7] *Instituto de Astrofisica de Canarias Via Lactea, s/n E38205 – La Laguna (Tenerife), Spain*
[8] *Blagoveschensk State Pedagogical University, Lenin Str, 104, Blagoveschensk 675000, Russia*
[9] *Skobeltsyn Institute of Nuclear Physics, Lomonosov MSU, 1, Leninskie gory, Moscow 119991, Russia*
[10] *Kislovodsk Solar Station of the Pulkovo Observatory, P.O. Box 45, ul. Gagarina 100, Kislovodsk 357700, Russia*

## ABSTRACT
We report on MASTER optical observations of an afterglow-like optical and X-ray transient AT2021lfa/ZTF21aayokph. We
detected the initial steady brightening of the transient at $7\sigma$ confidence level. This allowed us to use smooth optical self-similar
emission of GRBs model to constrain the explosion time to better than 14 min as well as to estimate its initial Lorentz factor
$\Gamma_0 = 20 \pm 10$. Taking into consideration the low $\Gamma_0$ and non-detection in gamma-rays, we classify this transient as the first failed
GRB afterglow.

## 1. INTRODUCTION

The long gamma-ray bursts (GRBs; Klebesadel et al. 1973; Mazets
et al. 1974) are the most powerful electromagnetic bursts in the
Universe and are thought to be associated with the formation of
rapidly rotating black holes. Rapid rotation hinders the core from
the fast descent into the event horizon and provides enough time to
turn a significant part of the gravitational and rotational energy into
electromagnetic emission. Due to the magnetic field and rotation, this
energy powers two narrow relativistic jets (Katz & Piran 1997). The
emission from the collimated relativistic jets may result in a new type
of phenomenon – 'orphan' optical, radio, or X-ray transient with no
corresponding gamma-ray burst (Rhoads 1997; Huang et al. 2002;
Nakar & Piran 2003). Over last 20 yr, numerous attempts have been
made to discover such transients to no avail (Levinson et al. 2002;
Robert et al. 2002; Rykoff et al. 2005; Gal-Yam et al. 2006; Rau et
al. 2006; Malacrino et al. 2007; Cenko et al. 2013; Ho et al. 2018,
2020; Law et al. 2018; Huang et al. 2020).

In this paper, we report MASTER-OAFA (Lipunov et al. 2010,
2022) optical observations of an afterglow-like optical and X-ray
transient AT2021lfa/ZTF21aayokph (Lipunov et al. 2021; Yao et al.
2021a), discuss its temporal behavior and constrain on the explosion

time calculated with the smooth optical self-similar emission of
GRBs model.

## 2. OBSERVATIONS

At 05:34:48 UTC 2021 May 4, the Zwicky Transient Facility de-
tected an unusual optical transient (OT) AT2021lfa/ZTF21aayokph
at $r = 18.6 \pm 0.08$ and 1.9 h later at $g = 18.8 \pm 0.11$, with no
detection 2 d earlier (Yao et al. 2021b). Follow-up observations
made by ZTF (Masci et al. 2019), Liverpool Telescope (Steele et
al. 2004), DDOTI (Watson et al. 2016, 2021), RATIR (Butler et al.
2012,2021), *Swift*-XRT, Gemini South, Mondy and Koshka, Lowell
Discovery Telescope, Assy and LBT (Fu et al. 2021; Ho et al. 2021;
Kim et al. 2021; O'Connor et al. 2021; Pankov et al. 2021; Rossi
et al. 2021; Yao et al. 2021b) revealed its steady fading, redshift of
$z = 1.063$, host galaxy and an X-ray transient with a luminosity
typical of an X-ray afterglow at the epoch.

Three hours before ZTF, during the routine MASTER global
robotic net sky survey (Lipunov et al. 2010, 2022), MASTER-
OAFA robotic telescope had detected the OT in six unfiltered frames
(Lipunov et al. 2021). The light curve (Fig. 1 and Table 1) shows
the steady brightening of the OT, with the significance of $7\sigma$, starting
from the first frame. The OT magnitude during the brightening
phase was lower than the ZTF first detection magnitude. Hence, we
conclude that peak brightness was located between the last MASTER
frame and the first ZTF detection.

★ E-mail: lipunov@sai.msu.ru

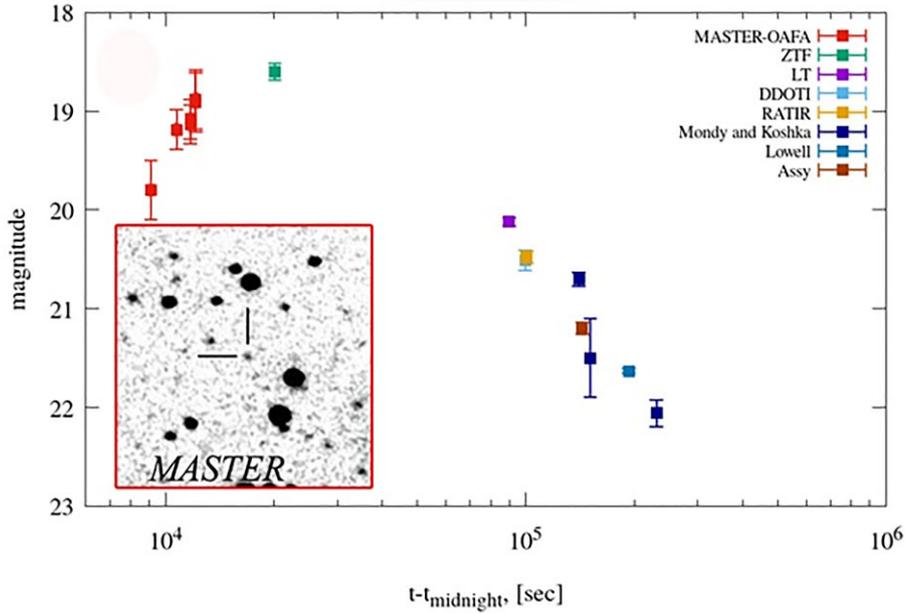

**Figure 1.** The optical light curve of MASTER OT J123248.62−012924.5 (AT2021lfa/ZTF21aayokph) combining MASTER (Table 1), ZTF, Liverpool Telescope, DDOTI, RATIR, Lowell, Mondy, Assy, Koshka (see details in Observations section) data, error bars are given at confidence $1\sigma$ level, $t_{midnight}$ = 2021–05–04 00:00:00 UTC. The inset shows stacked MASTER-OAFA image of made on the 2021–05–04 with 1080 s exposure time.

**Table 1.** MASTER photometry of AT2021lfa.

| Time (UTC) | Exp (s) | Unfiltered magnitude | Error of magnitude |
|---|---|---|---|
| 2021–05–04 02:31:49 | 180 | 19.8 | 0.3 |
| 2021–05–04 02:58:31 | 180 | 19.2 | 0.2 |
| 2021–05–04 03:15:19 | 180 | 19.1 | 0.2 |
| 2021–05–04 03:21:59 | 180 | 19.1 | 0.2 |
| 2021–05–04 03:42:01 | 180 | 18.9 | 0.3 |
| 2021–05–04 04:05:38 | 180 | 18.9 | 0.3 |

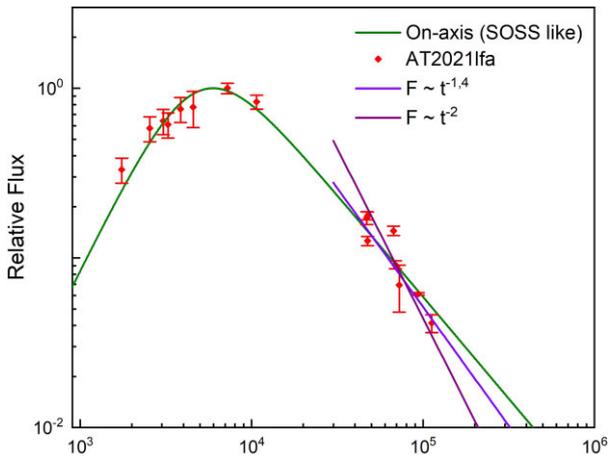

**Figure 2.** The SOSS model fit to the AT2021lfa light curve (green solid line). SOSS curve parameters are obtained through the analysis of the whole of data, $F \sim t^{-1/4}$ is obtained through the analysis of the decaying part (least squares model used for both analyses, see Section 3). Error bars corresponds to $1\sigma$ s.e.m.

## 2.2. MASTER photometry

MASTER global robotic net (Lipunov et al. 2010, 2019a, 2022; Kornilov et al. 2012) of Lomonosov Moscow State University was developed to investigate high-energy astrophysics events, in particular, to discover optical counterpart of gamma-ray bursts, gravitational wave events, high-energy neutrino, FRB and other events (Aartsen et al. 2017; Abbott et al. 2016; Buckley et al. 2018; Lipunov et al. 2016a,b,2018a,b,c,2019a,b,2020,2022). In addition to the targeted transient search, MASTER detects new optical transients during routine sky survey, in particular orphan afterglow and GRB afterglows at the earliest stage (Gorbovskoy et al. 2016; Lipunov et al. 2016b,2017a,b,c,d,2018b,c,2021,2022; Troja et al. 2017; Sadovnichy et al. 2018; Laskar et al. 2019; Zimnukhov et al. 2019; Ershova et al. 2020; Jordana-Mitjans et al. 2020). MASTER consists of nine fully robotic ground-based telescopes distributed all over the Earth with identical scientific equipment, which provides capability to continuously observe targets in a single photometric system. Each MASTER site consists of a twin 40 cm telescope with 4–8 square degrees field of view. The robotic observatories include the hardware, weather, ephemerides control, automatic evening/morning calibration, and central planner. The key feature of the observatories is the custom build autodetection real-time reduction software to discover optical transients, which includes primary image reduction, astrometry and photometry, extraction of new optical sources, and notification distribution for observers. To detect an optical transient candidate, the software automatically identifies all optical sources on the image with the *Gaia* EDR3 catalogue, then cross-correlate them with the MASTER archive data to detect new sources. The MASTER OT J123248.62–012924.5 was detected during the continuous survey running in the absence of GCN alerts (Barthelmy et al. 1998). To detect a transient during the survey the MASTER system uses at least two spaced in time unfiltered images of a sky region taken during the same night with a spatial shift of 0.5 arcmin to exclude any artefacts at CCD that could be erroneously identified as optical transients. After the automatic detection, the transient is sent to the central server for the classifications and further analysis.

**Table 2.** Best-fitting parameters of the averaged spectrum. NH is an equivalent hydrogen column. $Tb_{abs}$ and $zTb_{abs}$ are the Tuebingen–Boulder interstellar medium absorption models. NH(Galactic) corresponds to $Tb_{abs}$ model, NH(instrinsic) corresponds to $zTb_{abs}$ model. $c_{flux}$ is a flux for convolution model. $z$ is redshift, $\alpha_{ph}$ is photon index, $K$ is a normalization constant.

| | |
|---|---|
| NH (intrinsic) | $7\ (+59,\ -7) \times 10^{21}\ \mathrm{cm}^{-2}$ |
| NH (Galactic) | $1.95 \times 10^{20}\ \mathrm{cm}^{-2}$ |
| $z$ of absorber | 1.063 |
| Photon index ($\alpha_{ph}$) | $2.01\ (+1.79,\ -0.71)$ |
| Flux (0.3–10 keV) (observed) | $1.5\ (+1.0,\ -0.7) \times 10^{-13}\ \mathrm{erg\,cm}^{-2}\,\mathrm{s}^{-1}$ |
| Flux (0.3–10 keV) (unabsorbed) | $2.05\ (+11.31,\ -0.67) \times^{10-13}\ \mathrm{erg\,cm}^{-2}\,\mathrm{s}^{-1}$ |
| $C_{flux}$ (observed) | $3.30 \times 10^{-11}\ \mathrm{erg\,cm}^{-2}\,\mathrm{ct}^{-1}$ |
| $C_{flux}$ (unabsorbed) | $4.38 \times 10^{-11}\ \mathrm{erg\,cm}^{-2}\,\mathrm{ct}^{-1}$ |
| $W$-stat (d.o.f.) | 39.17 (36) |
| Spectrum exposure time | 9.7 ks |

To perform the OT photometry, we used the following procedure. Flux of the transient is measured in the 6 arcsec aperture and calibrated against several reference stars with a similar colour and G magnitudes from *Gaia* EDR3 (Gaia collaboration, 2016, 2020; Riello et al. 2021). The source magnitude uncertainties are estimated using a simple equation: $\Delta m = \sqrt{\frac{\sum_{i=1}^{N}(\bar{m}_i - m_{ij})^2}{N}}$, where $\bar{m}_i$ is the magnitude of calibration star $i$ averaged over the course of observations, $m_{ij}$ is the magnitude of check star $i$ on frame $j$, and $N$ is the number of calibration stars.

### 2.3. Light-curve fitting

To approximate the light curve of AT2021lfa, we used the smooth optical self-similar emission model (SOSS) for GRB afterglows (Lipunov et al. 2017a). The SOSS light curve is described by the following relation:

$$F = F_{max}\left(\frac{\beta\,\tau^{\beta-1}}{1+(\beta-1)\,\tau^{\beta}}\right)^{\alpha},\qquad(1)$$

where $\tau = (t - t_{trig})/(t_{max} - t_{trig})$, $t_{trig}$ is the burst trigger time, $t_{max}$ is the peak time, and $F_{max}$ is the peak optical flux. Here, we assume $\alpha = 1.2$, $\beta = 2.71$ (Lipunov et al. 2017a).

For AT2021lfa, all three model parameters were estimated by the least-squares model fit of the photometric data published in the GCN and collected by MASTER-OAFA (see Fig. 2).

### 2.4. X-ray observations

We analysed *Swift-XRT* observations at 1 d after the detection of AT2021lfa (Yao et al. 2021a) available from the *Swift-XRT* data repository (https://www.swift.ac.uk/xrt_spectra/), using XSPEC software from the HEASOFT 6.28 software package (Evans 2009). Assuming the source redshift of $z = 1.063$, the XRT spectrum was fitted with the power-law model taking into account the host and Milky Way interstellar medium absorption. The full model consists of the following components available in XSPEC: $c_{flux} \times Tb_{abs} \times zTb_{abs} \times$ PL, where $c_{flux}$ is a convolution model to calculate the flux; $Tb_{abs}$ and $zTb_{abs}$ are the Tuebingen –Boulder interstellar medium absorption models [these models calculate the cross-section for X-ray absorption and use NH (equivalent hydrogen column) and $z$ as parameters] for Milky Way and host, respectively; and the PL is $K \times E^{-\alpha_{ph}}$, $\alpha_{ph}$ is photon index, $K$ is a normalization constant, $E$ is an energy of the photon. As a result, we obtained parameters of the X-ray transient spectrum (Table 2 and Fig. 3).

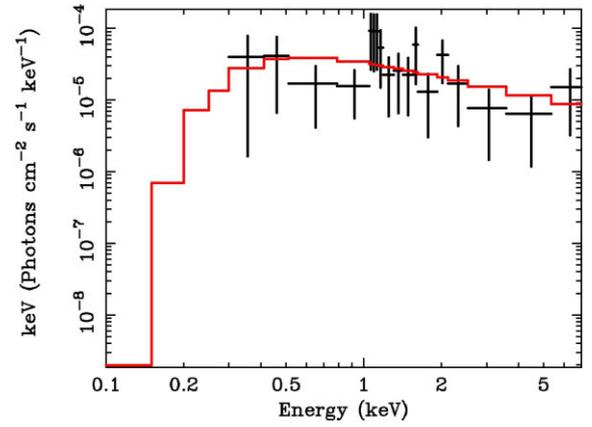

**Figure 3.** Averaged X-ray spectrum of AT2021lfa at 1 d. Error bars are at $1\sigma$ confidence level.

To estimate prompt isotropic peak gamma-ray luminosity, we used a set of correlations between X-ray flux and gamma-ray luminosity, derived for a large sample of GRBs (Dainotti et al. 2017), in particular, correlation between X-ray flux at 11 h after the burst in the rest frame and prompt gamma-ray fluence (D'Avanzo et al. 2012). *Swift-XRT* observations were conducted 28 h after the derived $t_{trig}$, assuming $z = 1.063$ rest-frame time of observations is 13.6 h, so $E_{\gamma,\ iso} > 10^{51}$ erg. This means that such a burst could have been detected and we checked the data from different space gamma observatories.

### 2.5. A gamma-ray counterpart search

We searched for an associated GRB in the estimated $t_{trig}$ interval in the data of the third Interplanetary Network (IPN), which consists of eight spacecrafts that provide all-sky full-time monitoring for GRBs. The most sensitive IPN instruments are the Swift Burst Alert

**Table 3.** The details of orbital instruments used for accompanying GRB search.

| Instrument | Energy band (keV) | Coverage comments |
|---|---|---|
| Wind (Konus) | 20–1500 | OT was visible during the entire interval, stable background |
| Fermi (GBM;NaI) | 8–1000 | OT was visible during the entire interval, variable background |
| INTEGRAL (SPI-ACS) | 75–8000 | OT was visible during the entire interval, stable background |
| *Swift* (BAT) | 15–350 | at $t_{\rm trig}$-840 s—$t_{\rm trig}$ −440 s *Swift* was in SAA; at $t_{\rm trig}$−440 s—$t_{\rm trig}$ + 840 s |
| | | OT was outside the BAT coded FoV |

**Table 4** *Gaia* reference stars used for the photometry.

| *Gaia* DR2 source ID | RA | Dec. | *G, m* |
|---|---|---|---|
| 3695332828608818048 | 188.20 392 | −1.46326 | 14.947 |
| 3695331935255619456 | 188.18 608 | −1.49577 | 14.569 |
| 3695331488579019648 | 188.19 002 | −1.50946 | 14.039 |
| 3695332038334845056 | 188.23 229 | −1.47246 | 16.586 |
| 3695328774159690112 | 188.25 280 | −1.50563 | 16.956 |
| 3695328464922044544 | 188.28 595 | −1.52887 | 14.879 |

Telescope (BAT; Barthelmy et al. 2005), the *Fermi* Gamma-ray Burst Monitor (GBM; Meegan et al. 2009), the INTEGRAL (von Kienlin et al. 2003) anticoincidence shield of the spectrometer SPI (Rau et al. 2005), and the Konus instrument on the Wind spacecraft (Aptekar et al. 1995). We searched for a GRB in the Swift-BAT GRB catalogue, the *Fermi*-GBM trigger list and subthreshold trigger list with reliability flag not equal to 2 and GCN archive (Barthelmy et al. 1998) and found no candidates.

The position of AT2021lfa was visible to GBM during the entire interval, the GBM showed highly variable count rate. *Swift* was passing the South Atlantic anomaly and was not collecting data for ∼20 per cent of the time, the source was outside the BAT coded field of view during the interval when the BAT was collecting data. By contrast, Konus-Wind and INTEGRAL (SPI-ACS) had complete coverage, in the 20–1500 keV and ∼80–8000 keV bands, respectively, and stable count rates (Table 3).

We found no significant (>5σ) excess over the background in the Konus-Wind waiting mode data on temporal scales from 2.944 s to 100 s. We also did not find any significant transients in the SPI-ACS data.

Using Konus-Wind waiting mode data, we estimate an upper limit (90 per cent confidence) on the peak flux, for a typical long GRB spectrum [the band function (Band et al. 1993) $\alpha_{\rm Band} = −1$, $\beta_{\rm Band} = −2.5$, and $E_{\rm p} = 300$ keV], to $3.4 \times 10^{-7}$ erg cm$^{-2}$ s$^{-1}$ (10–1500 keV, 2.944 s scale). A softer band spectrum, with $E_{\rm p} = 10$ keV, typical for low-luminosity GRBs results in a slightly lower limiting peak flux of $3.0 \times 10^{-7}$ erg cm$^{-2}$ s$^{-1}$.

For INTEGRAL SPI-ACS mean count rate at the time interval was 135 600 counts s$^{-1}$ in a bin of 50 ms. Using a conversion factor (Viganò & Mereghetti 2009), we derived an upper limit (90 per cent confidence) on the peak flux of $2.7 \times 10^{-7}$ erg cm$^{-2}$ s$^{-1}$.

## 3. SOSS EMISSION

From the OT light-curve shape (Fig. 1, Table 1 and reference stars, used for it, in Table 4), we conclude that peak brightness was located between the last MASTER frame and the first ZTF frame. Such a non-monotone smooth shape of the light curve hints towards AT2021lfa being an example of gamma-ray bursts with smooth optical self-similar emission (here on SOSS-like GRBs), firstly identified in 2017 (Lipunov et al. 2017a). SOSS-like GRB afterglows are described by a universal model that allows one to infer both burst time and peak time from the of the light curve. The model involves the synchrotron emission produced by the ultra-relativistic shock in the compressed ISM gas or in the stellar wind of the progenitor-star (Gruzinov & Waxman 1999). Decaying part of the light curve is defined by the cooling time of the relativistic electrons (adiabaticity of the shock) and the homogeneity of the environment. Such SOSS emission is virtually not connected to the prompt optical emission, and both can be seen at the same time.

Having fitted AT2021lfa's light curve with smooth optical self-similar emission model, we obtained following parameters: the explosion time $t_{\rm trig} = 2021–05–04$ 01:33 ± 00:14UTC; time of the light-curve peak $t_{\rm max} = 2021–05–04$ 04:52 ± 00:14 UTC; and peak magnitude $m_{\rm max} = 18.6 \pm 0.2$. Fig. 4 shows the light curve of the AT2021lfa compared with the brightest SOSS-like GRBs (including previously not published GRB180316A and GRB181213A, see photometry in Table 5). On the one hand, AT2021lfa does look similarly to them, but on the other hand, its peak time is the largest, which is most likely due to the selection effect: ZTF mostly surveys the sky in one day cadence, so long-duration events are more likely to be detected. One should also consider that despite a good fit by the SOSS light curve, the light curve decays as $t^{-1.4\pm0.2}$ rather than as $t^{-1.2}$ (asymptotic for SOSS-like GRBs) (see Fig. 2). This might be caused by a propagation of a forward shock in the progenitor's stellar wind in the slow cooling model (Zhang et al. 2006). This, however, does not reject the possibility of AT2021lfa having similar origin as GRBs.

AT2021lfa–GRB connection becomes even more apparent with the use of X-ray data from the Neil Gehrels Swift Observatory *Swift*-XRT (Ho et al. 2021). The flux in 0.3–10 keV flux was $2.9 \times 10^{-13}$ erg cm$^{-2}$ s$^{-1}$ d after ZTF detection that translates into luminosity of $1.9 \times 10^{45}$ erg s$^{-1}$ at $z = 1.063$ (Yao et al. 2021). Using this estimate, we can calculate the prompt gamma fluence of the transient if it was a GRB. Assuming that *Swift*-XRT observations were conducted approximately 1 d after the burst, we obtain following value $E_{\gamma, \rm iso} > 10^{51}$ erg. Comparison of X-ray spectrum of AT2021lfa with those of long GRBs shows little difference (Fig. 3, Table 2). Its photon index $\alpha_{\rm ph} = 2.01$ agrees with expected values, according to the standard afterglow theory (Sari 1997). Thus, we can assume that the observed X-ray transient resembles a gamma-ray burst in the late stages of the afterglow. Considering this, an estimation of gamma-ray fluence from X-ray detection as well as SOSS model we can assume that this is not a typical long GRB.

## 4. LORENTZ FACTOR ESTIMATION AND DISCUSSION

Currently, at least three models predict the existence of transients resembling GRB afterglows. These are on-axis orphan GRB (Nakar & Piran 2003), off-axis orphan GRB, or orphan afterglow (Rhoades 1997) and failed GRB or dirty fireball (Huang et al. 2002). The first one involves a jet of a typical GRB observed in the initial opening angle, but with no corresponding gamma-ray emission due to its higher collimation. Second model is similar, but jet is observed outside of an initial angle. Third one is based on a jet with high baryonic contamination due to which it has low Lorentz factor, much less than 100.

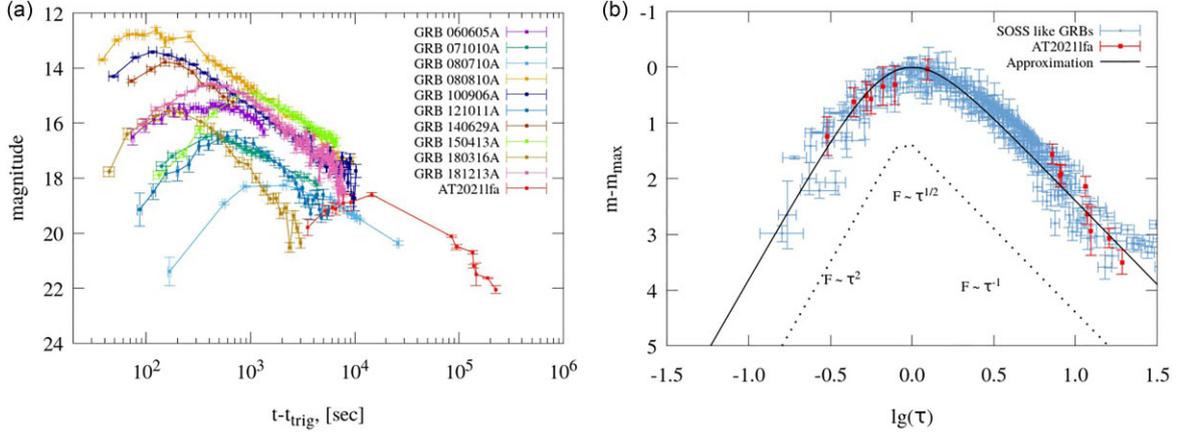

**Figure 4.** (a) Comparison of AT2021lfa with a SOSS-like (Lipunov et al. 2017a) afterglows for GRB 180316A and GRB181213A (Table 5). (b) AT2021lfa as a SOSS-like afterglow. Error bars are $1\sigma$ s.e.m.

To determine which type of orphan bursts the AT2021lfa belongs to, we turn to the calculations of the light curves of anisotropic jets moving in a homogeneous environment. Detailed studies show that at the initial stage (before the maximum is reached), off-axis afterglow light curves are rising significantly faster than in the case of on-axis jet (Huang et al. 2002; Granot et al. 2018; Xie & MacFadyen 2019). The difference is significant even at small viewing angles. Comparing the behavior of the SOSS curve with the simulated off-axis jet light curve (Fig. 5), we conclude that AT2021lfa is not an off-axis orphan GRB. On-axis orphan GRBs supposedly have similar light curves to failed GRBs but have a much higher initial Lorentz factor.

We can estimate it using two methods. First is a simple model of fireball expansion with an external shock wave (Sari 1997). Light curve in the standard afterglow model reaches maximum when mass of a material swept up from the circumburst medium by an external shock approximately equals mass of an ejected material.

$$\frac{E_0}{\Gamma_0^2 c^2} \approx \frac{4}{3}\pi m_p n R_d^3. \tag{2}$$

It occurs at a time $t_{p,z} \approx R_d/2\Gamma_0^2 c$ at a distance $R_d$ from the centre of the explosion, where $t_{p,z} = \frac{t_p}{1+z}$ and $t_p = t_{max} - t_{trig} = (12.0 \pm 0.8)\,10^3$s. By excluding the deceleration radius $R_d$ from the last two expressions, the initial value of the Lorentz factor is easily estimated as

$$\Gamma_0 \approx \left(\frac{3}{32\pi c^5 m_p}\right)^{1/8} \left(\frac{E_0}{n}\right)^{1/8} (t_{p,z})^{-3/8} \propto \left(\frac{E_0}{n}\right)^{1/8} (t_{p,z})^{-3/8}. \tag{3}$$

In principle, similar optical light curve can be produced during the propagation of a reverse shock wave. However, estimate of the shock wave Lorentz factor does not change (Hascoet et al. 2014) and detection of X-ray emission favours an forward one.

It should be noted that steepness of the optical light curve is more consistent with the radiation of a shock wave travelling in the stellar wind. Estimates of the initial Lorentz factor will obviously change. Indeed, assuming that the density distribution in the stellar wind is described as $\rho = Ar^{-2}$, ratio (2) is rewritten as $E_0/\Gamma_0^2 c^2 \approx$

$16\pi A R_d$. Again excluding $R_d$, one can obtain

$$\Gamma_0 \approx 52 \left(\frac{E_{52}}{A_*}\right)^{1/4} (t_{p,z})^{-1/4}. \tag{4}$$

Here, $A_* = A/(5 * 10^{11})\,\text{g cm}^{-1}$ is typical wind parameter, $E_{52} = E/10^{52}$. Using upper limit of luminosity and $A_* = 1$, one can obtain $\Gamma_0 \approx 10$.

It is also possible to estimate $\Gamma_0$ using a set of correlations between parameters of the light curve and an optical peak, Lorentz factor and gamma-ray fluence of GRBs (En-Wei Liang et al. 2010). From these correlations, we obtained estimates of isotropic gamma-ray fluence and initial Lorentz factor: fluence ranges from $10^{47}$ to $10^{51}$ erg which is consistent with the non-detection by Konus-Wind and INTEGRAL SPI-ACS and initial Lorentz factor $\Gamma_0 = (13.5 - 30.9)$ .

To verify correctness of these correlations in the case of AT2021lfa, we used correlations between temporal parameters that we obtained from SOSS light curve due to a lack of details on an actual light curve (Table 6):

$$\log t_d = (0.48 \pm 0.13) + (1.06 \pm 0.06)\log t_r, \tag{5}$$

$$\log t_d = (-0.09 \pm 0.29) + (1.17 \pm 0.11)\log t_p, \tag{6}$$

$$\log t_r = (-0.54 \pm 0.22) + (1.11 \pm 0.08)\log t_p, \tag{7}$$

$$\log \omega = (0.05 \pm 0.27) + (1.16 \pm 0.10)\log t_p. \tag{8}$$

One can see that the parameters satisfy them and do not deviate by more than $1\sigma$. We thus may assume that correlations show true picture despite some of them are inconsistent, mainly correlations between peak optical luminosity and isotropic gamma fluence and between cosmic-frame peak time $t_{p,z}$ and isotropic gamma fluence; estimate of isotropic gamma fluence ranges from $10^{51}$ erg in the former to $10^{47} - 10^{48}$ erg in the latter. They, however, both indicate towards non-detection of AT2021lfa by Konus-Wind and INTEGRAL SPI-ACS as they place an upper limit on the isotropic gamma fluence at $10^{52}$ erg.

Since we established correctness of these correlations, we can estimate initial Lorentz factor $\Gamma_0$ using rest-frame peak time $t_{p,z} = 5800 \pm 400$ s:

$$\log \Gamma_0 = (3.69 \pm 0.09) - (0.63 \pm 0.04)\log t_{p,z} = 1.31 \pm 0.48, \tag{9}$$

**Table 5.** MASTER Photometry data of GRB180316A and GRB181213A.

| $t-t_{trig}$ (s) | Exposition (s) | Magnitude | Error |
|---|---|---|---|
| GRB180316A | | | |
| 44 | 10 | 17.77 | 0.18 |
| 66 | 10 | 16.40 | 0.20 |
| 92 | 20 | 16.04 | 0.17 |
| 124 | 20 | 15.75 | 0.18 |
| 163 | 30 | 15.56 | 0.18 |
| 211 | 40 | 15.63 | 0.18 |
| 334 | 50 | 15.95 | 0.17 |
| 416 | 60 | 16.20 | 0.18 |
| 514 | 80 | 16.48 | 0.18 |
| 631 | 90 | 17.03 | 0.18 |
| 774 | 120 | 17.42 | 0.18 |
| 942 | 140 | 17.50 | 0.18 |
| 1129 | 170 | 18.00 | 0.18 |
| 1338 | 180 | 18.45 | 0.18 |
| 1547 | 180 | 18.76 | 0.18 |
| 1754 | 180 | 18.73 | 0.18 |
| 1962 | 180 | 19.29 | 0.18 |
| 2169 | 180 | 19.09 | 0.18 |
| 2379 | 180 | 20.52 | 0.18 |
| 2587 | 180 | 19.36 | 0.18 |
| 2793 | 180 | 19.82 | 0.18 |
| 3007 | 180 | 20.36 | 0.18 |
| GRB181213A | | | |
| 91 | 20 | 16.06 | 0.05 |
| 125 | 30 | 15.48 | 0.06 |
| 165 | 30 | 15.30 | 0.02 |
| 214 | 40 | 15.00 | 0.02 |
| 273 | 50 | 14.85 | 0.02 |
| 342 | 70 | 14.65 | 0.01 |
| 428 | 90 | 14.57 | 0.02 |
| 513 | 100 | 14.70 | 0.01 |
| 598 | 120 | 14.65 | 0.02 |
| 686 | 140 | 14.83 | 0.01 |
| 773 | 150 | 14.88 | 0.02 |
| 858 | 170 | 14.94 | 0.01 |
| 946 | 180 | 15.06 | 0.02 |
| 1031 | 180 | 15.18 | 0.02 |
| 1143 | 180 | 15.32 | 0.02 |
| 1229 | 180 | 15.43 | 0.02 |
| 1316 | 180 | 15.51 | 0.04 |
| 1402 | 180 | 15.54 | 0.02 |
| 1488 | 180 | 15.51 | 0.02 |
| 1576 | 180 | 15.73 | 0.02 |
| 1663 | 180 | 15.92 | 0.03 |
| 1751 | 180 | 15.84 | 0.03 |
| 1837 | 180 | 16.18 | 0.05 |
| 1923 | 180 | 16.01 | 0.03 |
| 2008 | 180 | 16.18 | 0.05 |
| 2093 | 180 | 16.32 | 0.04 |
| 2178 | 180 | 16.46 | 0.11 |
| 2266 | 180 | 16.15 | 0.06 |
| 2352 | 180 | 16.23 | 0.05 |
| 2439 | 180 | 16.33 | 0.07 |
| 2525 | 180 | 16.43 | 0.03 |
| 2611 | 180 | 16.29 | 0.05 |
| 2697 | 180 | 16.94 | 0.15 |
| 2868 | 180 | 16.61 | 0.04 |
| 2955 | 180 | 16.59 | 0.13 |
| 3040 | 180 | 16.72 | 0.06 |
| 3129 | 180 | 16.90 | 0.07 |
| 3216 | 180 | 16.93 | 0.11 |
| 3302 | 180 | 16.56 | 0.05 |
| 3388 | 180 | 16.36 | 0.09 |

**Table 5** – *continued*

| $t-t_{trig}$ (s) | Exposition (s) | Magnitude | Error |
|---|---|---|---|
| 3474 | 180 | 16.96 | 0.15 |
| 3561 | 180 | 16.63 | 0.06 |
| 3647 | 180 | 16.93 | 0.04 |
| 3733 | 180 | 17.50 | 0.06 |
| 3819 | 180 | 17.24 | 0.05 |
| 3906 | 180 | 17.14 | 0.05 |
| 3991 | 180 | 16.46 | 0.06 |
| 4200 | 180 | 17.18 | 0.07 |
| 4285 | 180 | 17.17 | 0.13 |
| 4372 | 180 | 17.17 | 0.05 |
| 4458 | 180 | 16.98 | 0.10 |
| 4545 | 180 | 18.09 | 0.07 |
| 4631 | 180 | 17.60 | 0.06 |
| 4719 | 180 | 17.08 | 0.14 |
| 4805 | 180 | 17.10 | 0.11 |
| 4892 | 180 | 17.44 | 0.06 |
| 4981 | 180 | 16.90 | 0.10 |
| 5067 | 180 | 17.29 | 0.05 |
| 5328 | 180 | 17.27 | 0.05 |
| 5414 | 180 | 16.85 | 0.09 |
| 5500 | 180 | 17.36 | 0.17 |
| 5588 | 180 | 17.24 | 0.17 |
| 5676 | 180 | 17.94 | 0.07 |
| 5762 | 180 | 17.41 | 0.06 |
| 5847 | 180 | 17.39 | 0.07 |
| 5933 | 180 | 17.19 | 0.08 |
| 6018 | 180 | 17.75 | 0.06 |
| 6106 | 180 | 17.21 | 0.15 |
| 6192 | 180 | 17.17 | 0.05 |
| 6279 | 180 | 17.59 | 0.06 |
| 6365 | 180 | 17.65 | 0.06 |
| 6569 | 180 | 17.73 | 0.06 |
| 6656 | 180 | 18.24 | 0.08 |
| 6742 | 180 | 17.45 | 0.06 |
| 6853 | 180 | 18.48 | 0.08 |
| 6940 | 180 | 18.90 | 0.10 |
| 7026 | 180 | 18.18 | 0.08 |
| 7111 | 180 | 17.93 | 0.07 |
| 7198 | 180 | 18.07 | 0.07 |
| 7286 | 180 | 18.48 | 0.08 |
| 7458 | 180 | 17.61 | 0.06 |
| 7545 | 180 | 17.72 | 0.06 |
| 7632 | 180 | 18.38 | 0.08 |
| 7717 | 180 | 18.61 | 0.09 |
| 7804 | 180 | 18.46 | 0.08 |

or $\Gamma_0 = (13.5 - 30.9)$. This value of Lorentz factor does not contradict non-detection by gamma-ray telescopes due to another correlation: $\Gamma_0 = 195 \times E_{\gamma,iso,52}^{0.27}$ that corresponds to $E_{\gamma,iso} = 10^{49}$ erg.

As both methods give Lorentz factor much less than 100, with the latter confirming non-detection by gamma-ray telescopes due to transient's inherently low gamma-ray fluence, we identify AT2021lfaas a failed GRB or dirty fireball.

## 5. RESULTS AND CONCLUSIONS

We have shown that the optical transient AT2021lfa/MASTER OT J123248.62−012924.5 is an orphan burst with a GRB-like X-ray afterglow without the presence of the gamma-ray burst itself. This phenomenon can be explained within the framework of the so-called dirty fireball model with a low Lorentz factor and an un-derdeveloped jet. Moreover, using the SOSS-emission model, it was possible to find the moment of the explosion. Modelling AT2021lfa light curve with smooth optical self-similar emission model, we obtained following values: moment of the explosion $t_{trig} = 2021$–05–04 01:33 ± 00:14UTC; moment of the light-curve peak $t_{max} = 2021$–05–04 04:52 ± 00:14 UTC; peak magnitude $m_{max} = 18.6 \pm 0.2$.

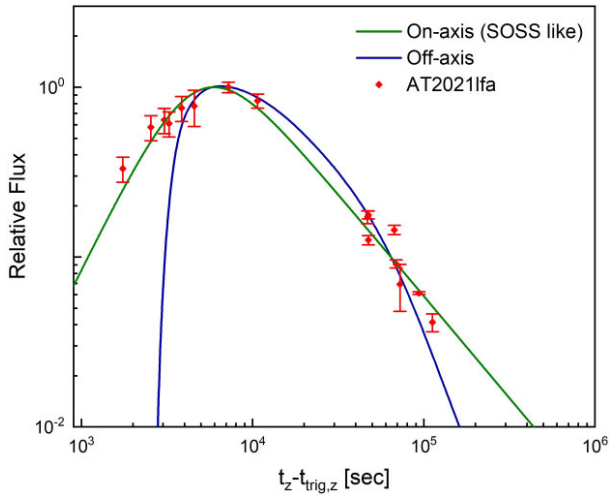

**Figure 5.** On-axis and off-axis (viewing angle 0.4 rad) SOSS-like orphan GRB afterglow in comparison to AT2021lfa. Initial half-opening angle for both cases $\theta_{init} = 0.2$, viewing angle for an off-axis case $\theta_{view} = 0.4$. Off-axis light curve is simulated using a top-hat Blandford-McKee model (Blandford & McKee 1976, 2019). Error bars are at $1\sigma$ confidence level.

**Table 6.** Temporal parameters of AT2021lfa.

| | |
|---|---|
| $t_p$, s | 11 970 ± 840 |
| $\omega$, s | 27 240 ± 2530 |
| $t_r$, s | 5260 ± 360 |
| $t_d$, s | 32 500 ± 2500 |
| $\log t_p$ | 4.08 ± 0.02 |
| $\log \omega$ | 4.43 ± 0,09 |
| $\log t_r$ | 3.72 ± 0.07 |
| $\log t_d$ | 4.51 ± 0.08 |

Of course, SOSS type emission of GRB does not necessarily state that it will be replicated in the exact way every time. In a similar way the afterglow shows monotonous power law behavior with a constant slope only with the averaged light curves, while in reality they can show some features such as optical flares, 'knees', etc. Initial stage of SOSS emission might be the disguised and 'broken' optical emission correlated with the gamma-ray emission that originates in the inner shockwave. It is obvious that heterogeneity in the density distribution of the progenitor-star's stellar wind connected with the variations in the physical parameters—outflow speed, magnetic field strength, can cause the optical emission to disappear. We also agree that the non-monotonic character of the prompt optical emission in GRB, found in this paper, is just an idealistic model of self-similar nature of the GRB ignition and transition to decaying. This behavior may be distorted by features concealing the true nature of the phenomenon.

## ACKNOWLEDGEMENTS

We dedicate this work to Viktor Kornilov, one of the key creators of the MASTER Global Network, who passed away in 2021 May. We acknowledge support by the Development programme of Lomonosov Moscow State University (MASTER equipment) and RFBR grant number 19–29-11011. NB work was performed using the UNU ≪Astrophysical Complex of MSU-ISU≫ facility (agreement 13.UNU.21.0007). The research is carried out using the equipment of the shared research facilities of Lomonosov MSU HPC computing resources (Sadovnichy et al. 2013). This work made use of data supplied by the UK Swift Science Data Centre at the University of Leicester. This work has made use of data from the European Space Agency (ESA) mission *Gaia*, processed by the Gaia Data Processing and Analysis Consortium (DPAC). Finally, we want to thank the anonymous referee for the comments, which have substantially improved the manuscript.

## DATA AVAILABILITY

The data presented in this work can be made available based on the individual request to the corresponding authors.